\documentclass[a4paper]{amsart}

\usepackage{amsmath}
\usepackage{amssymb}
\usepackage{amsfonts}
\usepackage{amsthm}

\newtheorem{lemma}{Lemma}[section]
\newtheorem{theorem}{Theorem}

\newtheorem{corollary}[lemma]{Corollary}
\theoremstyle{definition}

\theoremstyle{definition}
\newtheorem{remark}[lemma]{Remark}
\theoremstyle{definition}

{\catcode `\@=11 \global\let\AddToReset=\@addtoreset}
\AddToReset{equation}{section}

\newcommand{\gH}{\mathfrak{H}}
\newcommand{\gS}{\mathfrak{S}}

\newcommand{\gSto}{\mathfrak{S}_1^{P^0}}
\renewcommand{\S}{\mathcal{S}}
\newcommand{\E}{\mathcal{E}}

\newcommand{\h}{{\mathcal{H}}}
\newcommand{\cC}{\mathcal{C}}
\newcommand{\cH}{\mathcal{H}}
\newcommand{\C}{\mathbb{C}}
\newcommand{\R}{\mathbb{R}}

\newcommand{\vp}{\varphi}

\newcommand{\ie}{{\sl i.e.\/ }}
\newcommand{\cf}{{cf.\/ }}
\newcommand{\eg}{{\sl e.g.\ }}

\newcommand{\alp}{\boldsymbol{\alpha}}
\newcommand{\calc}{{\mathcal{C}}}
\newcommand{\sto}{{\gS_2(\gH_\Lambda)}}

\newcommand{\vpq}{{\varphi_Q}}
\newcommand\ii{{\ensuremath {\infty}}}

\newcommand{\norm}[1]{ \left| \! \left| #1 \right| \! \right| }

\def\tr{\mathop{\rm tr}\nolimits} 
\def\Tr{{\rm Tr}_{\C^4}}

\begin{document}

\title[Generalized Dirac-Fock type evolution equation]
{Existence of global-in-time solutions to a generalized Dirac-Fock type evolution equation}

\thanks{C.~H. and C.~S. acknowledge support from the Wittgenstein Award 2000 of 
Peter Markowich, to whom C.~H. is grateful for the warm hospitality at the 
Wolfgang Pauli Institute Vienna, where this work has been started. C.H. and
M.L. acknowledge support by  the European Union's IHP
network Analysis \& Quantum HPRN-CT-2002-00277. 
C.~S. has been supported by the APART grant of the Austrian Academy of Sciences.}

\author[C. Hainzl]{Christian Hainzl}
\address{University of Copenhagen, Department of Mathematics,
  Universitetsparken 5, 2100 Copenhagen} \email{hainzl@math.ku.dk}

\author[M. Lewin]{Mathieu Lewin}
 \address{University of Copenhagen, Department of Mathematics,
  Universitetsparken 5, 2100 Copenhagen} \email{lewin@math.ku.dk}

\author[C. Sparber]{Christof Sparber}
 \address{Department of Numerical Mathematics, University of M\"unster, Einsteinstra\ss e 62, 
 D-48149 M\"unster \& 
 Wolfgang Pauli Institute Vienna c/o Faculty of Mathematics, Vienna University, Nordbergstra\ss e 15, A-1090 Vienna, Austria}
\email{christof.sparber@univie.ac.at}


\begin{abstract}
We consider a generalized Dirac-Fock type evolution equation deduced from no-photon Quantum Electrodynamics, 
which describes the
self-consistent time-evolution of relativistic electrons, the observable ones as well as those filling up
the Dirac sea. This equation has been originally introduced by  Dirac in 1934 in a simplified form. 
Since we work in a Hartree-Fock type approximation,
the elements describing the physical state of the electrons are infinite rank projectors. Using the
Bogoliubov-Dirac-Fock formalism, introduced by Chaix-Iracane
({\it J. Phys. B.}, 22, 3791--3814, 1989), and recently established by
Hainzl-Lewin-S\'er\'e, we prove the existence of global-in-time
solutions of the considered evolution equation.
\end{abstract}

\date{Received 9 December 2004, revised 4 March 2005}

\maketitle

\section{Introduction}\label{s1}

The quantum mechanical description of relativistic 
spin-$1/2$ particles, say electrons, is based on the famous \emph{Dirac operator}, \ie 
$$
D^0:=-i\alp\cdot\nabla+\beta=-i\sum_{k=1}^3\alpha_k\partial_{x_k}+\beta,
$$
acting on $\mathfrak H=L^2(\R^3,\C^4)$,
where $\alp=(\alpha_1,\alpha_2,\alpha_3)$ and $\beta$ are the standard Dirac matrices \cite{Th}. Its main drawback, from
a physical as well as from a mathematical point of view, is the fact that its spectrum is not bounded from below, since
$$\sigma(D^0)=(-\ii,-1]\cup[1,\ii).$$
To circumvent this problem, Dirac \cite{D1,D2} postulated that the negative energy states are all filled
with one electron each, thereby taking into account Pauli's exclusion
principle. 
More precisely, within this framework the free vacuum state 
has to be considered as being an infinite Slater determinant 
$\Omega_0=\psi^0_1\wedge\psi^0_2\wedge\cdots\wedge\psi^0_i\wedge\cdots$, where $(\psi_i^0)_{i\geq1}$ 
is an orthonormal basis of the negative spectral subspace $\mathfrak H_-^0:=\chi_{(-\ii,0)}(D^0)L^2(\R^3,\C^4)$. 
The density matrix of $\Omega_0$ is then an infinite rank orthogonal projector 
given by 
$$P^0:=\chi_{(-\ii,0)}(D^0)=\sum_{i\geq1}|\psi_i^0\rangle\langle\psi_i^0|.
$$

This distribution of negative energy electrons is
postulated to be \emph{unobservable} on account of its uniformity. However, 
in the presence of an external potential, these so-called \emph{virtual} electrons react such that 
the vacuum becomes \emph{polarized}. The study of this polarization effects plays
an important role in modern Quantum Electrodynamics (QED), \cf \cite{GRS, GMR, Ue} and the 
references given in \cite{HLS1,HLS2}. In generality, Dirac in \cite{D1,D2}
approximates electrons, the observable ones as well as the ones
filling the sea, by an  orthogonal projector $P$ on $\mathfrak H$ with
infinite rank, 
which is interpreted as the density matrix of a formal infinite Slater
determinant 
$\Omega_P=\psi^P_1\wedge\psi^P_2\wedge\cdots\wedge\psi^P_i\wedge\cdots$,
where $(\psi_i^P)_{i\geq1}$ is an orthonormal basis of $PL^2(\R^3,\C^4)$. 

Our goal in this paper is to study a time-dependent equation describing the evolution of such an infinite rank orthogonal projector $P(t)$. 
This equation is obtained from the Bogoliubov-Dirac-Fock (BDF) model which is a mean-field approximation of \emph{no-photon} QED and 
has been introduced by Chaix-Iracane in \cite{CI}. It formally reads 
\begin{equation}
\label{peq_intro}
i \frac{d}{dt} P(t) = [D_{Q(t)}, P(t)],
\end{equation}
where $Q(t):=P(t)-P^0$ and $D_{Q(t)}$ is a \emph{mean-field operator} taking into account not only external fields, 
but also the \emph{self-consistent potentials} created by the state $P(t)$ itself:
\begin{equation}
D_Q:=D^0+\varphi+\alpha\rho_Q\ast\frac{1}{|\cdot|}-\alpha\frac{Q(x,y)}{|x-y|}.
\label{def_DQ_intro}
\end{equation} 
Here $\rho_Q(x)=\Tr(Q(x,x))=\Tr(P-P^0)(x,x)$ is the density of charge, $[\cdot,\cdot]$ denotes the usual commutator bracket, 
$\alpha$ is Sommerfeld's fine structure constant and $\varphi$ is some given external potential. We shall be mainly interested in the case where
$$
\varphi=-\alpha n\ast\frac{1}{|\cdot|}\, ,
$$
for some appropriately defined charge density $n$. Equation \eqref{peq_intro} is similar to 
the well known time-dependent Hartree-Fock model of non-relativistic quantum mechanics, see, \eg  \cite{CLB, CG}. 
The operator $Q(t)=P(t)-P^0$ is interpreted as the \emph{renormalized density matrix} (\ie measured with respect to the free vacuum $P^0$) 
of the quantum state described by $P(t)$. When no external field is present, \ie $\varphi=0$, the free vacuum $P^0$ is a
stationary solution of the equation \eqref{peq_intro}, since $Q(t)\equiv P^0-P^0=0$ and of course $[D^0,P^0]=0$. 

An equation very similar to \eqref{peq_intro} has been first introduced by Dirac himself in \cite{D1,D2}. 
Compared to equation \eqref{def_DQ_intro} though, he neglected the exchange term $-\alpha{Q(x,y)}/{|x-y|}$, 
\cf \cite[Eq. $(2)$]{D1}, \cite[Eq. $(3)$]{D2} and \cite[Eq. $(2.5.18)$]{QED}.
This last term of \eqref{peq_intro} is a consequence of Pauli's principle and is again classical in mean-field theories for fermions. 

The interpretation of \eqref{peq_intro}, in terms of the BDF model in no-photon QED, 
is roughly speaking as follows: in \cite{CI}, Chaix and Iracane consider the free relativistic Fock space in which 
they define a class of states furnished by \emph{Bogoliubov transformations} of the free vacuum. 
Each of these states can then be equivalently represented by its density matrix 
which is an orthogonal projector $P$, such that $Q=P-P^0$ is Hilbert-Schmidt. 
In \cite{CI}, the energy of such a BDF state $\Omega_P$ is computed and proved to be only depending on $Q=P-P^0$. 
Details can also be found in the appendix of \cite{HLS1}.
The equation \eqref{peq_intro} can be regarded as the natural evolution equation of the BDF model, 
since the right hand side of \eqref{peq_intro} is nothing but the derivative of the BDF energy under 
the constraint that $P=Q+P^0$ is an orthogonal projector. 
Our equation of interest \eqref{peq_intro} therefore describes the time-evolution of the 
associated state $\Omega_{P(t)}$, within the BDF class in Fock space.

In comparison to other relativistic theories, which usually do \emph{not} take into account the behavior of the vacuum, 
the BDF model of \cite{CI} admits the extremely important property that its associated energy 
is a \emph{bounded-below functional}, \cf \cite{BBHS,CIL,HLS1}. This property in the following enables 
us to prove the existence of global-in-time solutions to equation \eqref{peq_intro}. 

>From a mathematical point of view, solving the time-dependent non-linear equation \eqref{peq_intro} is not an easy task due to the fact that $P(t)$ is an infinite-rank projector. 
Indeed we shall heavily rely on the framework and results provided by 
Hainzl-Lewin-S\'er\'e in \cite{HLS1,HLS2}, where the existence of a minimizer $\bar P$ of the BDF energy is proved. 
This minimizer is found to be the solution of the self-consistent equation 
$$
\bar P=\chi_{(-\ii,0)}(D_{\bar Q}), \quad \mbox{ with $\bar Q=\bar P-P^0$,}
$$
and thus $\bar P$ comprises a stationary solution of \eqref{peq_intro}. 

Observe that the equation \eqref{peq_intro} includes the usual self-consistent 
interaction of real electrons, as well as additionally the interaction of real
electrons with the Dirac sea (virtual electrons) and its corresponding self-interaction. 
In the stationary case, one can minimize the BDF energy-functional, 
having fixed the total charge $N$ of $Q$. 
A minimizer $\tilde P$, with corresponding charge $N$, is then found to be the solution of 
an equation of the form (see \cite{CI,HLS3} and \cite[Remark 6]{HLS1})
\begin{equation}
\tilde P=\chi_{(-\ii,\lambda)}(D_{\tilde Q}), \quad \mbox{ with $\tilde Q=\tilde P-P^0$,}
\label{BDF_eq_mol}
\end{equation} 
where $\lambda$ is an Euler-Lagrange multiplier interpreted as a chemical potential. 
The projector $\tilde P$ is then also a stationary solution of our equation \eqref{peq_intro}. 
It can be naturally decomposed via 
$$\tilde P:=\tilde\Pi+\tilde\gamma$$
where $\tilde\gamma = \chi_{[0,\lambda)}(D_{\tilde Q}) = \sum_{j=1}^N|\phi_j\rangle\langle\phi_j|$ 
is a finite rank projector representing $N$ electrons, and $\tilde\Pi=\chi_{(-\ii,0)}(D_{\tilde Q})$ is an infinite rank 
projector, representing the Dirac sea.
Using \eqref{BDF_eq_mol}, it can easily be seen that the $\phi_j$'s are solutions of the Dirac-Fock equations \cite{ES1}, 
perturbed by the vacuum polarization potentials created by $\tilde\Pi$. 
Since these potentials are of higher order in the coupling constant $\alpha$, the usual Dirac-Fock model is 
therefore obtained as a non-variational approximation of the BDF model. 
However, for a general projector $P$, and in particular for our time-dependent solution $P(t)$, there is {\em a priori} no way of
performing the above given decomposition between real electrons and the Dirac sea.

As usual in QED, see, \eg \cite{BjDr}, we shall be forced to 
introduce an \emph{ultraviolet momentum cut-off} $\Lambda>0$, in order to obtain a well-defined local charge density $\rho_Q$. 
The only constraint on this cut-off $\Lambda$ will be its finiteness, which from a physical point of view corresponds 
to the assumption that electrons can \emph{not} shrink to a point. (Remark that $\Lambda \sim 10^{280} mc^2$ corresponds to
an electron radius of $10^{-280} cm$.) Such a point-like limit would cause rather unphysical effects, 
as discussed for example in \cite{HLS2} and already suggested earlier by Landau et al. \cite{La, LaPo}. 
In other words, we want to stress the fact that the UV cut-off 
is \emph{not} used as a mathematical tool in order to simplify the problem. Rather, the considered model 
\eqref{peq_intro} has no meaning if this cut-off is removed, simply because the function $\rho_Q$, appearing 
in \eqref{def_DQ_intro}, has no meaning. From a physical point of view, one may argue that all 
observables in the following are cut-off dependent and they diverge as $\Lambda\to\ii$. 
In the physics literature these divergencies are formally removed by means of a 
renormalization procedure \cite{BjDr, Dy}, which however is clearly beyond the 
scope of the present work. We thus choose of a \emph{finite}, but 
{\it arbitrarily} large cut-off $\Lambda$ and refer to \cite{HLS2} for comments and rigorous results 
concerning the behavior of the BDF model in the limit $\Lambda\to\ii$.

Finally note that our existence result is also valid if the exchange
term in \eqref{peq_intro} is dropped, leading to the original model considered by Dirac in \cite{D1,D2}. 

\medskip

The paper is now organized as follows: 
in the next section, we introduce the model and state our main result. Its proof is then provided in Section \ref{proofs}.

\section{Definitions and main results}\label{s2}

In this section we introduce the basic setting for our generalized Dirac-Fock type model
and state our main result. We follow \cite{HLS1,HLS2} and implement an 
ultraviolet cut-off $\Lambda$ by considering the following Hilbert space
$$ \gH_\Lambda = \{f \in  L^2(\R^3, \C^4)\ | \ \mathrm{supp}\hat f \subset B(0,\Lambda)\}.$$
The only restriction on $\Lambda$ will be its finiteness.
For any Hilbert-Schmidt operator 
\begin{equation}
Q\in \gS_2(\gH_\Lambda):=\left\{Q \ |\ \tr |Q|^2= \tr Q^*Q <\infty \right\},
\end{equation}
the associated charge-density $\rho_Q$ is a well-defined function in $L^2(\R^3)$. 
In the Fourier domain, it is given by \cite[Equ. (9)]{HLS1}
\begin{equation}
\label{def_rhoQ}
 \widehat{\rho_Q}(k) = \frac 1{(2\pi)^{3/2}} \int_{|p| \leq \Lambda } \Tr\left( \hat Q(p+k/2,p-k/2) \right)dp.
\end{equation} 
Notice that the function $Q\in\gS_2(\gH_\Lambda)\mapsto \rho_Q\in
L^2(\R^3)$ is continuous. Like in \cite{HLS2}, we 
introduce the following Hilbert space
$$\cC:=\left\{\rho\ |\ \int_{\R^3}\frac{|\widehat{\rho}(k)|^2}{|k|^2}dk<\ii\right\},$$
which is rather the Fourier inverse of $L^2(\R^3)$, weighted by $1/|k|^2$, and equipped with the so-called \emph{Coulomb norm}
$$
\norm{\rho}_{\cC}:=\left(\int_{\R^3}\frac{|\widehat{\rho}(k)|^2}{|k|^2}dk\right)^{1/2}.
$$
We consider a time-independent external field of the form
$$\varphi= - \alpha n\ast\frac{1}{|\cdot|},$$
where $n$ is some fixed (time-independent) function in $\cC$. 
Typically, one may think of $n$ as being the charge density created by some given system of smeared-out nuclei, 
with $\int_{\R^3}n(x)\,dx=Z$, denoting the total number of nuclei. 
For the following let us define our main Hilbert space \cite{HLS2}
$${\mathcal H}_\Lambda = \{  Q \in \gS_2(\gH_\Lambda), \, \, \rho_Q
\in \mathcal{C}\},$$ 
and its associated norm 
$$\|Q\|=\left(\norm{Q}_\sto^2 + \norm{\rho_Q}^2_{\mathcal{C}} \right)^{1/2} .$$
Moreover, let us denote by $\mathcal{P}_\Lambda$ the orthogonal projection of $L^2(\R^3)$ 
onto its closed subspace $\gH_\Lambda$. Then, for any $Q\in\cH_\Lambda$, 
\begin{equation}
\label{def_DQ}
D_Q:=\mathcal{P}_\Lambda\left(D^0+\alpha(\rho_Q-n)\ast\frac{1}{|\cdot|}-\alpha\frac{Q(x,y)}{|x-y|}\right)\mathcal{P}_\Lambda
\end{equation} 
is a well-defined self-adjoint bounded operator on $\gH_\Lambda$, \cf \cite[Lemma 4]{HLS1}. 
Notice that without the use of a projection $\mathcal{P}_\Lambda$ onto $\gH_\Lambda$, the operator 
$D_Q$ would not necessarily stabilize $\gH_\Lambda$. 
However, in the following we shall frequently omit to write the projection $\mathcal{P}_\Lambda$, for the sake of simplicity.

Our goal is then to study the following Cauchy problem
\begin{equation}\label{peq}
\left \{\begin{array}{l}
\begin{aligned}
i \frac{d}{dt} P(t) =& \, [D_{Q(t)}, P(t)],\\
P(0)= & \, P_I,\\
P(t)^2 = & \, P(t), \,\,\, Q(t) = P(t) - P^0 \in \cH_\Lambda,
\end{aligned}\\
\end{array} 
\right.
\end{equation}
 where
$$P^0=\chi_{(-\ii,0)}\left(\mathcal P_{\Lambda}D^0\mathcal P_{\Lambda}\right),$$
and $D_Q$ is defined in \eqref{def_DQ}. The operator $P_I=(P_I)^2$ is some
initially given orthogonal projector on $\gH_\Lambda$, which is such that $Q_I:=P_I-P^0\in\cH_\Lambda$. 
\begin{remark}
The requirement $P(t)-P^0\in\gS_2(\gH_\Lambda)$ implies, according to
the Shale-Stinespring Theorem \cite{KS,Rui, ShSt}, that the Fock
space representation associated with $P(t)$ is, for any $t$,
equivalent to the one associated with $P^0$ (see also \cite{Th} and, \eg,
\cite[Appendix]{HLS1}). Therefore $P(t)$ is associated with a BDF state in the free Fock space which is a Bogoliubov rotation of the free vacuum.
\end{remark}
Notice that we do \emph{not} assume $Q_I$ to be trace-class, since we strongly believe that this property will not be conserved
along the solutions, \ie $Q(t)\notin\gS_1(\gH_\Lambda)$ as soon as $t>0$, 
even if one starts with a $Q_I \in \gS_1(\gH_\Lambda)$. This belief
is supported by the fact that the stationary solutions constructed 
in \cite{HLS1, HLS2} are in general not trace-class. 
Of course, this forces us to add an UV cut-off $\Lambda$ in order to obtain a well-defined 
density via \eqref{def_rhoQ}. 

On the other hand, $Q(t)$ can be shown to be indeed $P^0$-trace class, a concept
introduced in \cite{HLS1} and which we redefine here for the reader's convenience: an operator $A\in \gS_2(\gH_\Lambda)$ is said to be 
\emph{$P^0$-trace class}, if $A^{++}:=(1-P^0)A(1-P^0)$ and $A^{--}:=P^0 A P^0$ are
trace-class ($\in \gS_1(\gH_\Lambda)$).
The associated $P^0$-trace of $A$ is then given by
\begin{equation}
\tr_{P^0} A = \tr A^{++} + \tr A^{--}.
\end{equation}
Notice, if $A$ is even trace-class then  
$$
\tr_{P^0} A = \tr A, \quad A\in \gS_1(\gH_\Lambda).
$$
In the following, we denote by
$\gSto(\gH_\Lambda)$ the set of all $P^0$-trace class operators and remark $\gSto(\gH_\Lambda)\subset\gS_2(\gH_\Lambda)$, 
by definition. 
It was shown in \cite{HLS1} that any difference of two projectors satisfying the \emph{Shale-Stinespring criterion},
\ie $Q=P-P^0 \in \gS_2(\gH_\Lambda)$, is automatically in $\gSto(\gH_\Lambda)$. This yields for 
the solution of \eqref{peq}, if it exists, that $P(t)$ will be such that
$$
Q(t)=P(t)-P^0\in\gSto(\gH_\Lambda).
$$
Moreover, by \cite[Lemma 2]{HLS1}, we know that $\tr_{P^0}(Q)$ is always an integer, and thus an appropriate 
tool for describing charge sectors. In the following, we shall show that {\it the total 
charge is a conserved quantity} along the solution, \ie 
$$
\tr_{P^0}(Q(t))=\tr_{P^0}(Q_I).$$

As we shall see, another conserved quantity for $Q(t)$ 
is the {\it Bogoliubov-Dirac-Fock energy}
\cite{CI,BBHS,HLS1,HLS2}. This energy is defined, for any
$Q=P-P^0\in\cH_\Lambda$ 
($P$ being a orthogonal projector), by 
\begin{equation}
\label{energy}
\mathcal{E}(Q)=\tr_{P^0}(D^0 Q)-\alpha D(\rho_Q,n) +\frac{\alpha}{2}D(\rho_Q, \rho_Q)
-\frac{\alpha}{2}\iint_{\R^6}
\frac{|Q(x,y)|^2}{|x-y|}\, dx\,dy,
\end{equation}
where
$$
D(f, g) = 4\pi\int_{\R^3}\frac{\overline{\widehat{f}(k)}\widehat{g}(k)}{|k|^2} \, dk.
$$ 
In \cite{HLS1,HLS2}, it is shown that $\mathcal{E}$ is bounded from below and moreover admits a
minimizer $\bar Q=\bar P-P^0$, which obeys the self-consistent equation
\begin{equation}\label{equfQ}
\bar P = \chi_{(-\infty,0)}\left(D_{\bar Q}\right).
\end{equation}
Obviously, this implies that $\bar P$ is a \emph{stationary solution} of our evolution equation \eqref{peq}, 
since $[D_{\bar Q},\bar P]=0$ holds. Notice that $\bar Q$ \emph{is never trace-class as soon as} $n\neq 0$. 

\medskip

We can now state the main result of this work:
\begin{theorem}\label{mt}
Let $\Lambda>0$, $0\leq \alpha< 4/\pi$ and $n \in {\mathcal C}$. 
Then, for any initial orthogonal projector $P_I$ such that $Q_I=P_I- P^0 \in \h_\Lambda$, there exists a unique maximal solution
$$P(t)\in C^1\left([0,\infty), P^0+\h_\Lambda\right)$$
of the Cauchy problem \eqref{peq}.
Moreover, one has
$$\tr_{P^0}(Q(t))=\tr_{P^0}(Q_I)\quad \text{and}\quad \mathcal{E}(Q(t))=\mathcal{E}(Q_I),$$
with $Q(t)=P(t)-P^0$, for all $t\in[0,\ii)$.
\end{theorem}
If the exchange term $Q(x,y)/|x-y|$ is neglected in $D_Q$, like in the original papers of Dirac \cite{D1,D2}, our 
theorem is valid for \emph{any} $\alpha\geq0$. This can be easily seen from the proof given in Section \ref{proofs} 
and the properties of the reduced BDF energy, as stated in \cite{HLS2}.
\begin{remark}
In the present work we adopt the original point of view of Dirac \cite{D1,D2} and others \cite{Hei,Wei} by fixing $P^0$ as the unobservable background. 
In principle other translation-invariant choices could be made.
Moreover Theorem \ref{mt} can be generalized, in a rather straightforward way, to cases in which 
a time-dependent external potential $\varphi(t)$, such that $|k|\hat
\varphi(t,k) \in C^1([0,\infty), \cC)$, is included. This situation might be considered in the study of 
time-adiabatic systems or in scattering theory.
\end{remark} 

Our method of proof is similar to classical arguments already used for the \emph{Hartree-Fock theory}, based on 
Schr\"odinger's equation, \cf \cite{CLB, CG} and the references given therein. 
We indeed solve the evolution-problem written in terms of $Q(t)=P(t)-P^0$, \ie
\begin{equation}\label{qeq}
\left \{
\begin{aligned}
i  \frac{d}{dt}  Q(t) = & \, [D_Q, Q] +[V_Q,P^0],\\
 Q(0)= & \, Q_I\in \cH_\Lambda,
\end{aligned}
\right.
\end{equation}
where 
$$V_Q = \mathcal{P}_\Lambda\left(\alpha \big(\rho_Q-n\big)\ast \frac{1}{|\cdot|} -\alpha \frac{Q(x,y)}{|x-y|}\right)\mathcal{P}_\Lambda.$$
We shall first prove the existence of a unique local-in-time solution in $\cH_\Lambda$ by classical arguments. 
Then, we show that the BDF energy \eqref{energy} is constant along this solutions, which consequently implies that it is indeed 
global-in-time, since $\E$ is bounded-below and coercive. Note that these are two fundamental properties not 
valid for the usual Dirac-Fock energy which is unbounded from below, each of its critical points 
having an infinite Morse index \cite{ES1,ES2}; see also \cite{EGS} for a related study.
\begin{remark} 
The existence of a bounded below energy-functional $\mathcal E$, conserved along the 
solution $P(t)$, is indeed a huge advantage of the BDF-model in
comparison to other nonlinear Dirac equations, see, \eg
\cite{EsVe, Ge, Gro, MNO}. There global-in-time solutions are obtained with quite strong restrictions on the considered initial data,
due to the non-existence of such an {\it a priori} energy-estimate. 
More precisely, sufficient smallness assumptions within appropriate Sobolev norms are required in \cite{EsVe, Ge, MNO}. 
Similar assumptions for initial data corresponding to sufficient small scattering states at $t=+\infty$ 
are used in the related study of \cite{FST,FST2}.
\end{remark}

\section{Proof of Theorem \ref{mt}}\label{proofs}
As mentioned above, we prove the existence of solutions to the system \eqref{qeq} written in terms of $Q(t)=P(t)-P^0$
\begin{equation}\label{qeq2}
\left \{
\begin{aligned}
i  \frac{d}{dt}  Q(t) = & \, F(Q(t)),\\
 Q(0)= & \, Q_I\in \cH_\Lambda,
\end{aligned}
\right.
\end{equation}
where 
\begin{equation}
F(Q)=[D_Q, Q] +[V_Q,P^0],\quad 
V_Q = \mathcal{P}_\Lambda\left(\alpha \big(\rho_Q-n\big)\ast \frac{1}{|\cdot|} -\alpha \frac{Q(x,y)}{|x-y|}\right)\mathcal{P}_\Lambda.
\label{def_F}
\end{equation} 

We start by proving first the existence of local-in-time solutions.
\begin{lemma}\label{local}
Let be $\Lambda>0$ and $n\in \cC$. Then, for any $Q_I  \in \h_\Lambda$, the nonlinear evolution
problem \eqref{qeq2} has a unique maximal solution
\begin{equation}
Q \in C^1\left([0,T), \h_\Lambda\right).
\end{equation}
Moreover, if the maximum time of existence $T$ is finite,
we have
\begin{equation}
\label{blow}
\lim_{t\nearrow T} \|Q(t)\| = \infty.
\end{equation}
\end{lemma} 
\begin{proof}
By Cauchy's Theorem, we need to show that $F$ is locally Lipschitz for
the $\cH_\Lambda$ topology. 
To this end, we first prove that there exist constants $C_1$, $C_2$ and $C_3$ such that
\begin{equation}
\left\{\begin{array}{l}
\norm{[D^\varphi,Q]}_\sto\leq C_1\|{Q}\|,\\
\norm{[V_Q,Q']}_\sto\leq C_2\|{Q}\| \,\|{Q'}\|,\\
\norm{[V_Q,P^0]}_\sto\leq C_3\|{Q}\|,
\end{array} \right.
\label{estimates}
\end{equation} 
where $D^\varphi:=\mathcal P_{\Lambda}(D^0-\alpha n\ast1/|\cdot|)\mathcal P_{\Lambda}$. 
Let us start by noting that for any $\rho\in\cC$, there exists a constant $\kappa$ such that
\begin{equation}
\norm{\rho\ast\frac{1}{|\cdot|}}_{\gS_{\ii}(\gH_\Lambda)}\leq \kappa E(\Lambda) \norm{\rho}_{\cC},
\label{estim_rho}
\end{equation} 
where $E(x)=\sqrt{1+x^2}$, and which is an immediate consequence of the following inequality proved in \cite[Proof of Theorem 1, step 3]{HLS2} 
$$\left|\rho\ast\frac{1}{|\cdot|}\right|\leq \kappa\norm{\rho}_\cC|D^0|.$$
Therefore, we can estimate
$$\norm{[D^\varphi, Q]}_\sto \leq 2
\norm{D^\varphi}_{\gS_\ii(\gH_\Lambda)}\norm{Q}_{\gS_2(\gH_\Lambda)}\leq
2(1+\alpha\kappa\norm{n}_\cC)E(\Lambda)\norm{Q}_\sto$$
and, denoting $\varphi_Q:=\alpha\rho_Q\ast1/|\cdot|$, we get
$$\norm{[\varphi_Q,Q']}_\sto \leq 2E(\Lambda)\alpha\kappa\norm{\rho_Q}_\cC \norm{Q'}_\sto.$$
On the other hand, $R_Q:={Q(x,y)}/{|x-y|}$ satisfies
$$\norm{R_Q}_\sto^2=\iint_{\R^6}\frac{|Q(x,y)|^2}{|x-y|^2}\,dx\,dy\leq
4 \tr\left((-\Delta)Q^2\right)\leq  4 \Lambda\norm{Q}_\sto^2,$$
by Hardy's inequality. 
Hence
$$\norm{[R_Q,P^0]}_\sto \leq 8\sqrt{\Lambda}\norm{Q}_\sto\norm{P^0}_{\gS_\ii(\gH_\Lambda)}\leq 8\sqrt{\Lambda}\norm{Q}_\sto,$$
and
$$\norm{[R_Q,Q']}_\sto \leq
2\norm{R_Q}_{\gS_\ii(\gH_\Lambda)}\norm{Q'}_\sto\leq 4 \sqrt{\Lambda}\norm{Q}_\sto\norm{Q'}_\sto.$$
To prove \eqref{estimates}, it therefore remains to estimate the Hilbert-Schmidt norm of $[\varphi_Q,P^0]$. 
Since the kernel of $\vp_QP^0$ in the Fourier domain is $(2\pi)^{3/2}\widehat{\vp_Q}(p-q)P^0(q)$, we obtain
\begin{eqnarray*}
\left|[\hat \vp_Q,\hat P^0](p,q)\right|^2 & = & (2\pi)^{3}|\widehat{\vp_Q}(p-q)|^2\Tr(P^0(p)-P^0(q))^2\\
 & = & 2(2\pi)^{3}|\widehat{\vp_Q}(p-q)|^2\Tr(P^0(p)P^0_\perp(q)),
\end{eqnarray*} 
where $P^0_\perp=1-P^0$.
By Lemma 12 in \cite{HLS1}, we have
\begin{equation}\label{Pineq}
\Tr(P^0(p)P^0_\perp(q))\leq \frac{|p-q|^2}{2E((p+q)/2)^2},
\end{equation}
and thus
$$\norm{[\vp_Q,P^0]}_\sto^2 \leq (2\pi)^{3}\iint_{|p|,|q|\leq\Lambda}\frac{|\widehat{\rho_Q}(p-q)|^2}{|p-q|^2E((p+q)/2)^2}\,dp\,dq,$$
which yields
$$ \norm{[\vp_Q, P^0]}_\sto\leq C_\Lambda\norm{\rho_Q}_\cC,$$
for some constant $C_\Lambda$ depending on the cut-off $\Lambda$. This consequently proves \eqref{estimates}.

It remains to study $F$ in the $\cC$ norm. To this end, we remark that the density corresponding to
$[\vpq,Q]$ vanishes, since
\begin{multline}
\hat \rho_{[\vpq,Q]}(k) = \iint \widehat{\vp_Q}(p+k/2-s) \Tr\widehat Q(s,p-k/2)\, dp\,ds \\ 
-\iint \widehat{\vp_Q}(s-p+k/2)\Tr\widehat Q(p+k/2,s)\, dp\,ds=0,
\end{multline}
by a simple variable transformation. Likewise the density of
$[\vpq,P^0]$ vanishes since the Dirac matrices $(\alpha_k)_{k=1}^3$ and $\beta$ have vanishing
$\C^4$-trace. Similarly the density of $[R_Q,Q]$ is identically zero 
due to the symmetry in $Q$. Therefore, it remains to prove the continuity of the linear terms 
$[\mathcal D^0,Q]$ and $[R_Q,P^0]$ in the $\cC$ norm.
We find 
\begin{eqnarray*}
\frac{\widehat{\rho_{[D^0,Q]}}(k)}{|k|} & = & \frac 1{(2\pi)^{3/2}|k|}\int_{|p| \leq \Lambda}\Tr\left(D^0(p+ k/2)\widehat Q(p+k/2,p-k/2)\right. \\ 
 & & \qquad \left. - \widehat Q(p+k/2, p - k/2) D^0(p - k/2)\right) dp \\
 & = &\frac 1{(2\pi)^{3/2}} \int_{|p|\leq \Lambda}\Tr\left( \frac {\alp \cdot k}{|k|} \widehat Q(p+ k/2,p- k/2) \right) dp
\end{eqnarray*}
and thus
$$\frac{|\widehat{\rho_{[D^0,Q]}}(k)|}{|k|}\leq C_0 |B(0,\Lambda)|^{1/2}
 \left( \int_{|p| \leq \Lambda} |\widehat Q|^2(p+  k/2, p-  k/2) dp\right)^{1/2},$$
with $C_0 = \sqrt{12}/(2\pi)^{3/2}$, using Schwarz's inequality for matrices as well as functions.
Squaring and integrating over $k$ implies
$$\norm{\rho_{[D^0,Q]}}_\calc\leq C_0 |B(0,\Lambda)|^{1/2}\norm{Q}_\sto.$$
Additionally 
\begin{multline}\label{rhoqp}
\frac{\hat \rho_{[R_Q,P^0]}(k)}{|k|} = \\ \int_{|p|\leq \Lambda} \Tr\left(\frac{1}{|k|}\left(\hat P^0(p+k/2) - \hat P^0(p-k/2)\right)
 R_Q(p+k/2,p-k/2)\right)dp.
\end{multline}
By inequality \eqref{Pineq} we thus have
$$\Tr\left((P^0(p+k/2) - P^0(p-k/2))^2\right)=2\Tr(P^0(p+k/2)P^0_\perp(p-k/2)) \leq  \frac{|k|^2}{E(p)^2},$$
which, using Schwarz's inequality first for matrices then in the integration over $p$, leads to
$$ \norm{\rho_{[R_Q,P^0]}}_\calc \leq C_\Lambda \norm{R_Q}_\sto\leq C_\Lambda \norm{Q}_\sto$$
and thus the assertion of Lemma \ref{local} is proved.
\end{proof} 

Let us now show that $P(t):=Q(t)+P^0$ is indeed a projection for any $t\in[0,T)$.
\begin{lemma}
Let be $\Lambda>0$, $n\in \cC$, $Q_I=P_I-P^0  \in \h_\Lambda$ and
$(Q(t),[0,T))$ given by Lemma \ref{local}. Then $P(t)=Q(t)+P^0$ is an
orthogonal 
projector.
\end{lemma} 
\begin{proof} 
Define a new unknown $A(t):=P(t)^2-P(t)$. Then $\dot A = \dot P P + P \dot P -\dot P$, which, together with \eqref{qeq}, 
yields  
\begin{equation}\label{P_is_proj}
\left \{
\begin{aligned}
i \frac{d}{dt} A(t) = & \, [D_{Q(t)}, A(t)],\\
A(0)= & \, P_I^2-P_I=0.
\end{aligned}
\right.
\end{equation}
Since $D_{Q(t)} \in C^1\left([0,T),\gS_\infty(\gH_\Lambda)\right)$  this
  equation obviously admits a unique maximal solution in
  $\gS_1(\gH_\Lambda)$ 
and therefore 
$A(t)=P(t)^2-P(t)=0$ for any $t\in[0,T)$.
\end{proof} 

As a consequence, we infer from Lemma 2 in \cite{HLS1} that $Q(t)=P(t) - P^0$ is $P^0$-trace-class and that its charge 
$\tr_{P^0}(Q(t))$ is an integer. Indeed, we also know that 
$$\tr_{P^0}(Q(t))=\tr(Q(t)^3).$$
Since $t\in[0,T)\mapsto Q(t)$ is continuous in the $\gS_2(\gH_\Lambda)$ topology and therefore also in the 
$\gS_3(\gH_\Lambda)$ topology, we obtain that $t\in[0,T)\mapsto \tr_{P^0}(Q(t))$ is continuous and hence constant, \ie
$$\forall t\in[0,T),\ \tr_{P^0}(Q(t))=\tr_{P^0}(Q_I).$$

\medskip

We now show that the BDF energy is also constant along the solution which will imply that $T=\ii$.
\begin{lemma}\label{energy_ct}
Let be $\Lambda>0$, $n\in \cC$, $Q_I=P_I-P^0  \in \h_\Lambda$ and $(Q(t),[0,T))$ given by Lemma \ref{local}. Then 
$\E(Q(t))=\E(Q_I)$ for any $t\in[0,T)$.
\end{lemma} 
\begin{proof}
Let us first show that $\dot{Q}(t)\in\gS_1^{P^0}(\gH_\Lambda)$ for any $t\in[0,T)$. Indeed $[V_Q,P^0]$ is obviously in $\gS_1^{P^0}(\gH_\Lambda)$ since 
$$P^0[V_Q,P^0]P^0=P^0_\perp[V_Q,P^0]P^0_\perp=0.$$
To proceed further, we denote $P_t:=\chi_{(-\ii,0)}( D_{Q(t)})$, where $D_{Q(t)}$ is defined as in \eqref{def_DQ}, \ie 
including the spectral projections $\mathcal P_{\Lambda}$. Then $P_t$ satisfies $P_t-P^0\in\gS_2(\gH_\Lambda)$, 
for any $t\in[0,T)$ by the results given in \cite{KS}. 

Let us now recall the following fact, proved in \cite[Lemma 1]{HLS1}, which we shall use: 
if $P$ and $P'$ are two orthogonal projectors, such that $P-P' \in \gS_2(\gH_{\Lambda})$, then it holds:
\begin{equation}\label{PPeq}
A \in \gS_1^P(\gH_{\Lambda}),\ \text{if and only if} \ A\in \gS_1^{P'}(\gH_{\Lambda}),\ 
\text{and $\tr_{P} A =\tr_{P'} A$.}
\end{equation}

Using \eqref{PPeq}, we deduce that $Q(t)\in \gS_1^{P_t}(\gH_\Lambda)=\gS_1^{P^0}(\gH_\Lambda)$ for any $t\in[0,T)$ and therefore
$$P_t[D_{Q(t)},Q(t)]P_t=[D_{Q(t)},P_tQ(t)P_t]\in\gS_1(\gH_\Lambda),$$
since $P_tQ(t)P_t\in\gS_1(\gH_\Lambda)$ and $D_{Q(t)}\in\gS_\ii(\gH_\Lambda)$. 
Using again \eqref{PPeq}, we infer $[D_{Q(t)},Q(t)]\in \gS_1^{P_t}(\gH_\Lambda)=\gS_1^{P^0}(\gH_\Lambda)$ and therefore $\dot Q(t)\in \gS_1^{P^0}(\gH_\Lambda)$, 
which easily implies
$$\frac{d}{dt}\tr_{P^0}(Q(t))=\tr_{P^0}(\dot{Q}(t)).$$
By means of Lemma 5 in \cite{HLS1}, that is 
$$
\tr_{P^0}(D^0 \dot Q(t))+\alpha D(\rho_Q-n,\rho_{\dot Q(t)}) - \alpha \tr \left(\frac{Q(x,y)}{|x-y|} \,\dot Q(t) \right)=
\tr_{P^0}(D_{Q(t)} \dot Q(t)),
$$
we consequently obtain
$$
\frac d{dt} \E(Q(t)) = \tr_{P^0}(D_{Q(t)}\dot Q(t)).
$$
Hence, inserting equation \eqref{qeq} gives
\begin{equation}
\frac d{dt} \E(Q(t))= -i \tr_{P^0}\left(D_{Q(t)}[D_{Q(t)},Q(t)]\right) -i \tr_{P^0}\left(D_{Q(t)}[V_{Q(t)},P^0]\right).
\label{der_energy}
\end{equation} 
The first term on the right hand side vanishes again due to \eqref{PPeq},
since $$\tr_{P^0}\left(D_{Q(t)}[D_{Q(t)},Q(t)]\right) = \tr_{P_t}\left(D_{Q(t)}[D_{Q(t)},Q(t)]\right) =0.$$
Concerning the second term on the right hand side of \eqref{der_energy}, we first notice 
that obviously $\tr_{P^0}(D^0[V_{Q(t)},P^0])=0$. 
Furthermore we evaluate
$$P^0V_Q[V_Q,P^0]P^0 = P^0V_Q P_\perp^0 V_Q P^0,$$
as well as $$P_\perp^0V_Q[V_Q,P^0]P_\perp^0 =  -P_\perp^0V_Q P^0 V_Q P_\perp^0.$$
Together with the fact that $P_\perp^0 V_Q P^0 \in \gS_2(\gH_\Lambda)$, this yields
$$\tr_{P^0}(V_Q[V_Q,P^0]) = \tr \left([P^0V_QP^0_\perp,P^0_\perp V_Q P^0]\right) = 0,$$
and thus it holds 
$$
\frac d{dt} \E(Q(t))=0,
$$
which concludes the proof.
\end{proof} 

This finally allows us to end the proof of Theorem \ref{mt}.
\begin{corollary}
Assume that $0\leq\alpha<4/\pi$. Let be $\Lambda>0$, $n\in \cC$,
$Q_I=P_I-P^0  \in \h_\Lambda$ and $(Q(t),[0,T))$ given by Lemma
\ref{local}. 
Then $T=\ii$.
\end{corollary} 
\begin{proof}
Since $P(t)$ is a projector, we have $-P^0 \leq Q(t) \leq 1-P^0$. We
know from \cite[Equ. (26)]{HLS2} that $\E$ is coercive, for $0\leq\alpha<4/\pi$, on the set 
$$
S_\Lambda= \{ Q \in \cH_\Lambda, \, -P^0\leq Q \leq 1-P^0 \},
$$ 
due to the following inequality 
$$
\mathcal E(Q(t))+\frac{\alpha}{2} \norm{n}^2_{\mathcal{C}}\geq \left(1-\alpha \frac{\pi}{4}\right)\tr_{P^0}(D^0Q(t))+
\frac{\alpha}{2}\norm{\rho_{Q(t)}-n}^2_{\mathcal{C}},
$$
and the fact that, when $Q\in\S_{\Lambda}$, it holds
$$
\tr_{P^0}(D^0Q)\geq \tr(|D^0|Q^2)\geq \norm{Q}_{\gS_2(\gH_\Lambda)}^2.
$$
Since $\mathcal E(Q(t))=\mathcal E(Q_I)$, this consequently implies that $\|Q(t)\|$ stays bounded  for $\alpha < 4/\pi$.
Therefore the maximum time of existence 
is $T =\infty$ and  Theorem \ref{mt} is proved.
\end{proof} 


\bibliographystyle{amsplain}

\begin{thebibliography}{99}

\bibitem{BBHS} V. Bach, J.~M. Barbaroux, B. Helffer, and H. Siedentop, \emph{On the Stability of the relativistic
electron-positron field}, Comm. Math. Phys. \textbf{201} (1999), 445--460.
\bibitem{BjDr} B.~J. Bjorken and S.~D. Drell, \emph{Relativistic quantum fields}, McGraw-Hill, New York, 1965. 
\bibitem{CI} P. Chaix and D. Iracane \emph{From quantum electrodynamics to mean field theory:
I. The Bogoliubov-Dirac-Fock formalism}, J. Phys. B. \textbf{22} (1989), 3791--3814.
\bibitem{CIL} P. Chaix, D. Iracane, and P.L. Lions, \emph{From quantum electrodynamics to mean field theory: II.
Variational stability of the vacuum of quantum electrodynamics in the mean-field approximation}, J. Phys. B.
\textbf{22} (1989), 3815--3828.
\bibitem{CLB} \'E. Canc\`es and C. Le Bris. {\it On the time-dependent Hartree-Fock equations coupled with a 
classical nuclear dynamics}, Math. Models Methods Appl. Sci. {\bf 9} (1999), no. 7, 963--990.
\bibitem{CG} J.M. Chadam and R.T Glassey. {\it Global existence of solutions to the Cauchy problem for 
time-dependent Hartree equations}, J. Math. Phys. \textbf{16} (1975), 1122--1230.
\bibitem{D1} P.~A.~M. Dirac, \emph{Th{\'e}orie du positron}, Solvay
  report, 203--212. Paris: Gauthier-Villars. {XXV} (1934), 353.
(reprinted in {\it Selected papers on Quantum Electrodynamics}, edited by J. Schwinger, Dover, 1958).
\bibitem{D2} P.~A.~M. Dirac, \emph{Discussion of the infinite distribution of electrons in the theory of the positron},
Proc. Camb. Philos. Soc. \textbf{30} (1934), 150--163.
\bibitem{Dy} F.~J. Dyson, \emph{The S Matrix in Quantum Electrodynamics}, Phys. Rev. {\bf 75}(11) (1949), 1736--1755. 
\bibitem{EsVe} M. Escobedo and L. Vega, \emph{A Semilinear Dirac Equation in $H^s(\R^3)$ for $s > 1$}, 
SIAM J. Math. Anal. {\bf 28} (1997), no. 2, 338--362.
\bibitem {ES1} M. Esteban and E. S\'er\'e, \emph{Solutions of the Dirac-Fock equations for atoms and molecules}, 
Comm. Math. Phys. {\bf 203} (1999), no. 3, 499--530.
\bibitem {ES2} M. Esteban and E. S\'er\'e, \emph{Nonrelativistic limit of the Dirac-Fock equations}, Ann. Henri Poincar\'e {\bf 2}
(2001), no. 5, 941--961.
\bibitem {EGS} M. Esteban, V. Georgiev, and E. S\'er\'e, \emph{Stationary solutions of the Maxwell-Dirac and the 
Klein-Gordon-Dirac equations}, Calc. Var. Part. Diff. Equ. {\bf 4} (1996), no. 3, 256--281. 
\bibitem{FST} M. Flato, J. Simon, C.~H. Taflin, \emph{On global solutions of the Maxwell-Dirac equations}, 
Comm. Math. Phys. \textbf{112} (1987), no. 1, 21--46.
\bibitem{FST2} M. Flato, J. Simon, C.~H. Taflin, {\it Asymptotic completeness, global existence and the 
infrared problem for the Maxwell-Dirac equations}, Mem. Amer. Math. Soc. {\bf 127} (1997), no. 606, x+311 pp.
\bibitem {Ge} V. Georgiev, \emph{Small amplitude solutions of the Maxwell-Dirac equations}, Indiana Univ.
Math. J. \textbf{40} (1991), no. 3, 845--883.
\bibitem{GRS} R. Glauber, W. Rarita, and P. Schwed, {\it Vacuum polarization effects on energy levels in {$\mu$}-mesonic atoms}, 
Phys. Rev. {\bf 120} (1960), no. 2, 609--613.
\bibitem{GMR} W.~Greiner, B.~M{\"u}ller, and J.~Rafelski, {\it Quantum Electrodynamics of Strong Fields}. Texts and Mongraphs in Physics, Springer-Verlag, 1985.
\bibitem {Gro} L. Gross, \emph{The Cauchy problem for the coupled Maxwell and Dirac equations},
Comm. Pure Appl. Math. \textbf{19} (1966), 1--15.
\bibitem{H} C. Hainzl, \emph{On the Vacuum Polarization Density caused by an External Field}, Ann. Henri Poincar\'e \textbf{5} (2004), 1137--1157.
\bibitem{HLS1} C. Hainzl, M. Lewin, and E. S\'er\'e, \emph{Existence of a stable polarized vacuum in the Bogoliubov-Dirac-Fock
approximation}, { Comm. Math. Phys.} {\it to appear}.
\bibitem{HLS2} C. Hainzl, M. Lewin, and E. S\'er\'e, \emph{Self-consistent solution for the polarized vacuum in a no-photon QED model}, { J. Phys. A: Math \& Gen.} {\it to appear}.
\bibitem{HLS3} C. Hainzl, M. Lewin, and E. S\'er\'e, {\it in preparation}.
\bibitem{HS} C. Hainzl, H. Siedentop, \emph{Non-Perturbative Mass and Charge Renormalization in Relativistic no-photon
Quantum Electrodynamics}, Comm. Math. Phys. {\bf 243} (2003), 241--260.
\bibitem{Hei} W. Heisenberg, \emph{Bemerkungen zur Diracschen Theorie des Positrons}, Zeits. f. Physik {\bf 90} (1934), 209--223.
\bibitem{KS} M. Klaus and G. Scharf, \emph{The regular external field problem in quantum electrodynamics}, Helv. Phys. Acta {\bf 50} (1977), 779-802.
\bibitem{La} L.~D. Landau, \emph{On the quantum theory of fields}, Pergamon Press, Oxford 1955. Reprinted in \emph{Collected 
papers of L.~D. Landau}, edited by D. Ter Haar, Pergamon Press, 1965.
\bibitem{LaPo} L.~D. Landau and I. Pomeranchuk, \emph{On point interactions in quantum electrodynamics}, Dokl. Akad. Nauk. SSSR {\bf 102}, 489--492. 
Reprinted in \emph{Collected papers of L.~D. Landau}, edited by D. Ter Haar, Pergamon Press, 1965.
\bibitem{MNO} S. Machihara, K. Nakanishi, and T. Ozawa, \emph{Small global solutions and the nonrelativistic limit 
for the nonlinear Dirac equation} Rev. Mat. Iberoam. {\bf 19} (2003), no. 1, 179--194. 
\bibitem{Rui} S.~N.~M. Ruijsenaars, {\it On Bogoliubov transformations for systems of relativistic charged particles},
J. Math. Phys. {\bf 18} (1977), no.3, 517--526, 1977.
\bibitem{QED} S. S. Schweber, {\it QED and the men who made it: Dyson, Feynman, Schwinger and Tomonaga}, Princeton University Press, 1994.
\bibitem {ShSt} D. Shale and W. Stinespring, \emph{Spinor representation of infinite orthogonal groups},
J. Math. and Mech. {\bf 14} (1965), 315-324.
\bibitem{Simon} B. Simon. {\it Trace Ideals and their Applications}. 
Vol  35 of {\it London Mathematical Society Lecture Notes Series}. 
Cambridge University Press, 1979.
\bibitem{Th} B. Thaller, {\it The Dirac Equation}, Springer Verlag, 1992.
\bibitem{Ue} E.A. Uehling, \emph{Polarization effects in the positron theory}, Phys. Rev. II. Ser. {\bf 48} (1935), 55--63.
\bibitem{Wei} V.~Weisskopf, {\em{\"U}ber die {E}lektrodynamik des {V}akuums auf {G}rund der {Q}uantentheorie des {E}lektrons}, {Math.-Fys. Medd., Danske Vid. Selsk.} {\bf 16} (1936), no. 3, 1--39. (reprinted in {\it Selected papers on Quantum Electrodynamics}, edited by J. Schwinger, Dover, 1958).
\end{thebibliography}

\end{document}